\documentclass[aps,graphics,epsf]{revtex4}
\begin{document}
\def\tablerule{\noalign {\vskip3truept\hrule\vskip3truept}}
\def\half{{1\over 2}}
\def \e {\eta }
\def \ee {({\eta\over \eta _0})}
\def \D {\mbox{D}}
\def\curl {\mbox{curl}\,}
\def \ep {\varepsilon}
\def \lleq {\lower0.9ex\hbox{ $\buildrel < \over \sim$} ~}
\def \ggeq {\lower0.9ex\hbox{ $\buildrel > \over \sim$} ~}
\newcommand{\sq}{\lower.25ex\hbox{\large$\Box$}}
\def \l {\Lambda}
\def\beq{\begin{equation}}
\def\eeq{\end{equation}}
\def\ber{\begin{eqnarray}}
\def\eer{\end{eqnarray}}
\def \ie {{\em i.e.~~}}
\def \apl {ApJ, }
\def \aps {ApJS, }
\def \pd {Phys. Rev. D, }
\def \prl {Phys. Rev. Lett., }
\def \pl {Phys. Lett., }
\def \np {Nucl. Phys., }

\title{An Exact Inflationary Solution On The Brane}
\author{M. Sami}
\email[Email: ]{sami@jamia.net}
\affiliation{Department of Physics, \\
Jamia Millia Islamia, New Delhi-110025, INDIA}

\begin{abstract}
    
    We study the evolution of universe with a single scalar field of constant potential
    minimally coupled
     to gravity in the brane world cosmology.We find an exact inflationary solution
     which is not in slow roll.We discuss the limiting cases of the solution.We show that
     at late times the solution is asymptotic to the de Sitter solution 
     independently of the 
     brane tension. For $t\rightarrow 0$ the solution leads to singularity but the nature of
     the approach to singularity depends upon the brane tension.
     
\end{abstract}
\maketitle

\section{Introduction}
 The higher dimensional cosmological models inspired by string models are under
 active investigation at present [ 1,2 ]. According to the brane World scenario 
 our four dimensional space time is realized as a boundary ( brane ) of higher
  dimensional space time ( the bulk ).The simplest possibility uses the embedding
   into five dimensional manifold [ 3 ]. In these models 
 the Einstein theory of a five dimensional space time is considered with a four
 dimensional boundary (3 brane ) on which the boundary conditions should be treated
 dynamically.The metric describing the full 4+1 dimensional space
 is not flat along the extra dimension ; it constitutes a slice of Anti-de Sitter space (AdS).
In this picture all the matter fields are confined to the bran while  gravity can propagate
in the bulk.
The small value of the true five dimensional Planck's mass in these models
is related to its large effective four dimensional value by the extremely large wrap of
five dimensional space .And this offers a possibility for alleviating the hierarchy problem
in the particle theory [ 1 ].  The field equations on the brane get contribution from dynamics
 of extra dimension and differ from the usual Einstein equations.This may have 
 important implications for cosmology.For instance , the Friedman equation gets
 modified by a term quadratic in density which enhances the prospects of 
 inflation [ 4, 6 ]. A particular attention has also been paid on the quantum creation of the brane world [ 5 ].

The equations of inflation are normally solved in slow roll
approximation. In this approximation the slow roll parameters
$\epsilon,\eta << 1$ [ 7 ]. The  parameters $\epsilon$ and $\eta$ are
related to the slope and the curvature of the field potential  respectively.The
slow roll conditions in the usual Friedman Cosmology as well as their
generalization for the brane world express the necessary but not sufficient conditions for
neglecting the acceleration term in comparison of the force and the friction
terms in the field equation.It is possible to find a solution of the field
equation which satisfies the slow roll conditions but has still a large
speed. One such solution in the usual 4-dimensional space time was given
in reference [ 8 ]. In this paper we obtain a similar solution on the brane.\par
 In the five dimensional brane world scenario the Friedman constraint
 equation acquires the generalized form ,

 \begin{equation}
 H^2={\frac{8\pi}{3M_4^2}}\rho(1+\rho/2\lambda_b)+{\Lambda_4 \over 3}+{E \over a^4(t)}
 \end{equation}
 where E is a constant which describes the bulk graviton effect on a 3-brane and
 $\Lambda_4$ is the four dimensional Planck's constant.The brane tension $\lambda_b$
 relates the four dimensional Planck's mass with its five dimensional counterpart
 as,
 \begin{equation}
 M_4=\sqrt{\frac{3}{4\pi}}(M_5^2/\sqrt\lambda_b)M_5
\end{equation} 
For simplicity we set $\Lambda_4$ equal to zero and also drop the last term as
otherwise the inflation would render it so , leading to the expression ,
\begin{equation}
H^2=\frac{8\pi}{3M_4^2}\rho(1+\rho/2\lambda_b)
\end{equation}
where $\rho={1 \over 2} \dot{\phi}^2+V(\phi)$ , if one is dealing with a universe 
dominated by a single scalar field minimally coupled to gravity. The equation of
motion of the field propagating on the brane is
\begin{equation}
\ddot{\phi}+3H \dot{\phi}+{dV \over d\phi}=0
\end{equation}
 \section{The Slow Roll Evolution and the Equation of State on the Brane}
The field equation and the Friedman equation on brane in slow roll approximation
have the form ,
\begin{equation}
3H\dot{\phi}=-{dV(\phi) \over d\phi)}
\end{equation}
\begin{equation}
H^2={\frac{8\pi}{3M_4^2}}V(1+V/2\lambda_b)
\end{equation} 
The adiabatic index in slow roll assumes the form ,
$$\gamma={\dot{\phi}^2/2-V\over\dot{\phi}^2/2+V}+1 \simeq\dot{\phi}^2/V$$
Using Eq. ( 5 ) we find 
\begin{equation}
\gamma=\left({2\over3}\right){M_4^2\over 16\pi}{1\over(1+V/2\lambda_b)}\left({V_{,\phi} \over V}\right)^2
\end{equation}
For inflation to occur in usual four dimensional space time,
\begin{equation}
\epsilon= {3\over2}{\gamma_4} < 1
\end{equation}
where $\gamma_4$ is the adiabatic index in the four dimensional space time
and $\epsilon$ is the slow roll parameter in the usual case.
However,the condition on $\gamma$ for inflation gets modified on brane.This can be seen
from the following.The equation for acceleration on brane takes the form ,
\begin{equation}
{\ddot{a}(t) \over a(t)}=-\left({8 \pi \over 6M_4^2}\right)\rho\left((1+3\omega)+\rho/\lambda_b(2+3\omega)\right)
\end{equation}
The condition for inflation on brane follows ,
\begin{equation}
\omega<-{1\over3}\left(1+2\rho/\lambda_b\right)/(1+\rho/\lambda_b)
\end{equation}
where $$\omega=\gamma-1$$
Or
\begin{equation}
\gamma <{2\over 3}(1+\rho/2\lambda_b)/(1+\rho/\lambda_b)
\end{equation}
Similar to usual 4-dimensional case we can again define the slow roll parameter,
\begin{equation}
\epsilon={3\over 2}\gamma_B
\end{equation}
where,
\begin{equation}
\gamma_B=\gamma{(1+\rho/\lambda_b)\over(1+\rho/2\lambda_b)}.
\end{equation}
$\gamma_B$ satisfies the same condition for inflation as in usual four 
dimensional space time and this leads to the correct expression for slow roll 
parameter [ 6 ],
\begin{equation}
\epsilon ={M_4^2\over 16 \pi}\left({V_{,\phi} \over V}\right)^2 \left[({1+V/\lambda_b)/(1+V/2\lambda_b)^2}\right]
\end{equation}

The state equation on brane takes the form  ,
\begin{equation}
\omega_B=\gamma_B-1
\end{equation}
and
\begin{equation}
\omega_B\equiv \omega (1+\rho/\lambda_b)/(1+\rho/2\lambda_b)
\end{equation}
which is  clear from(13).
\section{An Exact Inflationary Solution on the Brane}
In slow roll approximation one assumes that $V(\phi) >>\dot{\phi}^2/2$ and the
inflaton motion is friction dominated.The acceleration term can then be neglected
in equation ( 4 ).A necessary but not sufficient condition for this to happen on
the brane is given by ,
\begin{equation}
\epsilon ={M_4^2\over 16 \pi}\left({V_{,\phi} \over V}\right)^2
{ \left[{(1+V/\lambda_b)\over(1+V/2\lambda_b)^2}\right]}
<<1
\end{equation}
An exact solution of field equations which satisfies ( 17 ) but which is not in slow
roll i.e $\dot{\phi}$ is not small was written by Faraoni in Friedman cosmology [ 8 ].
We find a similar exact solution on the 3-brane.We shall look for an exact
 solution of evolution equations for a constant potential ,
\begin{equation}
V(\phi)=V_0=Constant
\end{equation}
for $\dot{\phi}$ not equal to zero.In this case the field equation readily solves
to give $\dot{\phi}$,
\begin{equation}
\dot{\phi}={C \over a^3}
\end{equation}

where C is an integration constant.The Friedman equation with constant potential
given by ( 18 ) and the expression for the field velocity given by ( 19 ) assumes
 the form ,
\begin{equation}
{dy \over dt}=\beta \sqrt{\left(\alpha e^{-6y}+1\right)\left
(\delta_1+\delta \alpha e^{-6y}\right)}
\end{equation}
Where $\alpha={C^2 \over 2V_0}$, $y=\ln{a}$, $\beta=\left({8\pi V_0^2 \over 3 M_p^2}\right)^{1/2}$
,$\delta={V_0 \over 2\lambda_b}$ and $\delta_1=\delta+1$.Equation( 20 ) can immediately be integrated
to yield the expression for the scale factor,
\begin{eqnarray}
a(t)&=&\left[\left({\alpha(\delta_1^2+\delta^2) \over 4A\delta_1}-{\delta\alpha \over 2A}\right)
e^{6\sqrt{\delta_1}\beta t}\right.\nonumber\\
&+&\left.{\alpha A \over 4\delta_1}
e^{-6 \sqrt{\delta_1}\beta t}-{\alpha(\delta_1+\delta) \over 2\delta_1}\right]^{1/6}
\end{eqnarray}
Where the integration constant A is chosen such that a(t) goes to zero
for t going to zero.And this fixes the value of A,
\begin{equation}
A=1+2\delta+(2+2\delta)\sqrt{\delta/(1+\delta)}
\end{equation}
\section{Investigation of the Solution}
The solution given by equation ( 17 ) has interesting asymptotic forms.For t going to 
zero
the scale factor vanishes and all the physical quantities like pressure and field
energy density diverge.The solution crucially depends upon brane parameter 
$\delta$ which
contains the brane tension$\lambda_b$ and the approach to singularity depends upon
this parameter.On the other hand at late times $t \rightarrow \infty $ the solution
asymptotically approaches the de Sitter solution independently of $\delta$.This is not
 surprising as it is well known that de Sitter solution is a late time attractor in the
4-dimensional space time and the same seems to be true on the brane also. However,
a rigorous phase space analysis analogous to reference[ 9 ] is required to be carried out on the brane. It would
be interesting to consider the high and low energy limits of the solution ( 21 ).In the
low energy limit $\delta<<1$ i.e $V_0<<2\lambda_b$ , the usual 
Friedman cosmology is retrieved and the expression ( 21 ) assumes the form[ 8 ],
\begin{equation}
a(t)=\alpha^{1/6}\sinh^{1/3}(3\beta t)
\end{equation}
The integration constant A in this case turns out to be equal to one.The equation
of state in this case is given by,
\begin{equation}
\omega\equiv {p \over \rho}=1-2\tanh^2(3\beta t)
\end{equation}
The brane effects  are most pronounced in high energy limit $\delta \rightarrow \infty$.
In this limit the second term in the expression does not survive.Expanding the 
coefficient of exponential in the first term into Teller series in $1/\delta$ and keeping
the first order term we obtain,
\begin{equation}
a(t)=\alpha^{1/6}\left(e^{6\sqrt{\delta}\beta t}-1\right)^{1/6}
\end{equation}
With $A=4\delta$.The solution can easily be found out by directly integrating
the generalized Friedman equation ( 3 ) in high energy limit. The equation of state in this case turns out to be,
\begin{equation}
{ p \over \rho}=2e^{-6\sqrt{\delta}\beta t}-1
\end{equation}
In low energy limit for t going to zero $a(t)\propto t^{1/3}$ where as in high energy
limit $a(t)\propto t^{1/6}$ i.e the approach to singularity is slower than the usual 
big bang scenario.The equation of state similar to Friedman case ( Eq( 24 ) ) interpolates
between the stiff matter equation of state $p=\rho$ and the vacuum state equation
$p=-\rho$.The general solution ( 21 ) interpolates between the solutions given by
( 23 ) and ( 25 ) as $\delta$ varies between zero(low energy limit) and infinity (high
energy limit).\par
To conclude, we have studied the evolution of the universe with a single scalar field of constant potential on the brane. We have obtained an exact inflationary solution which
is asymptotic to the de Sitter solution at late times independently of the brane tension.
At $t=0$ , the solution exhibits singularity and the approach to singularity depends upon the brane tension.

\section{ Acknowledgments}
I am thankful to V. Sahni , P. Sharan and T. Qureshi for useful discussions. I am also thankful
to IUCAA for hospitality where this work was started.\\


\end{document}